\documentclass[apj]{emulateapj} 

\shortauthors{Kilic et al.} 
\shorttitle{A Dark Spot on a White Dwarf}

\begin{document} 
\title{A Dark Spot on a Massive White Dwarf\altaffilmark{*}}

\author{
Mukremin Kilic\altaffilmark{1}, 
Alexandros Gianninas\altaffilmark{1},
Keaton J. Bell\altaffilmark{2}, 
Brandon Curd\altaffilmark{1},
Warren R. Brown\altaffilmark{3},
J. J. Hermes\altaffilmark{4},
Patrick Dufour\altaffilmark{5},
John P. Wisniewski\altaffilmark{1},
D. E. Winget\altaffilmark{2},
K. I. Winget\altaffilmark{2}
}
\altaffiltext{*}{This work is based on observations obtained at the Gemini Observatory,
McDonald Observatory, and the Apache Point Observatory 3.5-meter telescope. The latter
is owned and operated by the Astrophysical Research Consortium. Gemini Observatory is operated by the
Association of Universities for Research in Astronomy, Inc., under a cooperative agreement
with the NSF on behalf of the Gemini partnership: the National Science Foundation
(United States), the National Research Council (Canada), CONICYT (Chile), the Australian
Research Council (Australia), Minist\'{e}rio da Ci\^{e}ncia, Tecnologia e Inova\c{c}\~{a}o
(Brazil) and Ministerio de Ciencia, Tecnolog\'{i}a e Innovaci\'{o}n Productiva (Argentina).}
\altaffiltext{1}{Homer L. Dodge Department of Physics and Astronomy, University of Oklahoma, 440 W. Brooks St., Norman, OK, 73019, USA}
\altaffiltext{2}{Department of Astronomy, University of Texas at Austin, Austin, TX 78712, USA}
\altaffiltext{3}{Smithsonian Astrophysical Observatory, 60 Garden St., Cambridge, MA 02138, USA}
\altaffiltext{4}{Department of Physics, University of Warwick, Coventry CV4 7AL, UK}
\altaffiltext{5}{Institut de recherche sur les exoplan\'etes (iREx), D\'epartement de Physique, Universit\'e de Montr\'eal, C.P. 6128, Succ. Centre-Ville, Montr\'eal, Qu\'ebec H3C 3J7, Canada}

\begin{abstract}

We present the serendipitous discovery of eclipse-like events around the
massive white dwarf SDSS J152934.98+292801.9 (hereafter J1529+2928). We selected J1529+2928
for time-series photometry based on its spectroscopic temperature and surface gravity,
which place it near the ZZ Ceti instability strip. Instead of pulsations, we
detect photometric dips from this white dwarf every 38 minutes. Follow-up optical
spectroscopy observations with Gemini reveal no significant radial velocity
variations, ruling out stellar and brown dwarf companions. A disintegrating planet
around this white dwarf cannot explain the observed light curves in different
filters. Given the short period, the source of the photometric dips
must be a dark spot that comes into view every 38 min due to the rotation of the white dwarf.
Our optical spectroscopy does not show any evidence of Zeeman splitting of the
Balmer lines, limiting the magnetic field strength to $B<70$ kG. 
Since up to 15\% of white dwarfs display kG magnetic fields, such eclipse-like
events should be common around white dwarfs. We discuss the potential implications of
this discovery on transient surveys targeting white dwarfs, like the K2 mission and
the Large Synoptic Survey Telescope.

\end{abstract}


\section{Introduction}

White dwarfs are commonly used as spectrophotometric standards since
the majority of them do not show any significant variations. The first pulsating
white dwarf was in fact discovered by \citet{landolt68} as part of his survey
for new standard stars. Since then $>200$ other DAV, DBV, and PG 1159 type variable
white dwarfs have been identified in a variety of surveys \citep[e.g.,][]{mukadam04}.

Other sources of variability for white dwarfs include the relativistic
beaming effect \citep{loeb03,zucker07}, ellipsoidal variations,
and eclipses. However, there are only a handful of white dwarfs known to display these
effects \citep[][and references therein]{shporer10,kilic11,parsons13,hermes14,hallakoun15}.
Eclipses can be due to (sub)stellar or planetary companions, and \citet{vanderburg15} present
the first candidate for a disintegrating planet around a white dwarf in the K2 mission
data.

An alternate source of an `eclipse-like' event on a white dwarf is a dark spot, or a starspot.
There are several examples of white dwarfs with strong ($>1$ MG) magnetic fields that display
starspots, but a strong magnetic field is not always required or observed
\citep{wickram00}. RE J0317-853 belongs to the former category,
where a $B=$ 340-450 MG field is detected along with 0.2 mag peak-to-peak photometric
variations due to a starspot that comes into view every 725 s due to the rotation of
the white dwarf \citep{barstow95,ferrario97}. \citet{gary13} find 2.7 h period,
low-amplitude (5 mma) optical variations on WD 2359-434, a magnetic white dwarf
with a relatively low field strength of $B= 3.1 \pm 0.4$ kG \citep{aznar04}.
GD 394 presents an unusual case where \citet{dupuis00} detect 25\% amplitude variations
in the extreme ultraviolet with a period of 1.15 d, but there is no evidence of a magnetic
field down to a limit of 12 kG. \citet{dupuis00} interpret the observed photometric variations
as evidence of surface abundance inhomogeneities. 

The Kepler mission and the ongoing K2 mission are revealing that a large fraction of white
dwarfs are variable. In addition to BOKS 53856, a magnetic white dwarf that displays 4.9\% amplitude variations
every 6.1 h \citep{holberg11}, \citet{maoz15} find low-amplitude (60-2000 ppm)
variations in half of the 14 white dwarfs observed as part of the original Kepler mission.
This is unprecedented, as it implies that a large fraction of white dwarfs do indeed vary on hour-to-day
timescales. The K2 mission has already discovered a disintegrating planet and a double white dwarf
system in the first field \citep{vanderburg15,hallakoun15}, and a significant fraction of the K2 white dwarfs
do also show low-amplitude variations (N. Hallakoun 2015, private communication). 
Understanding the source of these variations is essential for transient surveys that are targeting
large numbers of white dwarfs.

Here we present the serendipitious discovery of 38 min period optical photometric variations
in an isolated massive white dwarf. Section 2 describes our target selection and follow-up observations,
whereas Section 3 presents the results from our photometric and spectroscopic analysis. Section 4 visits three
potential explanations for the observed photometric dips and demonstrates that a dark spot
is the most likely source. We discuss the potential implications of this discovery on the K2 mission
and the LSST, and conclude in Section 5.

\section{Target Selection and Observations}

\subsection{J1529+2928}

We chose J1529+2928 for follow-up time-series photometry based on its temperature and surface
gravity measurements from its Sloan Digital Sky Survey (SDSS) spectrum. \citet{kleinman13} classify
it as a potentially magnetic DAH white dwarf based on this spectrum. However, there is no evidence of
Zeeman splitting of the Balmer lines in our higher quality Gemini data (see below). 
Using the 1D atmosphere models for
non-magnetic DA white dwarfs and the fitting procedures outlined in \citet{gianninas11}, we find
$T_{\rm eff} = 11450$ K and $\log{g} = 8.88$ based on the SDSS data. These place J1529+2928 near
the empirical boundaries of the ZZ Ceti instability strip \citep{gianninas11}. J1529+2928 is slightly
cooler than the massive pulsating white dwarf BPM 37093, which has $T_{\rm eff} = 11920 \pm 190$ K,
$\log{g} = 8.81 \pm 0.05$, and pulsation periods between 512 and 635 s \citep{metcalfe04}. 

\subsection{Time-Series Photometry}

We acquired high speed photometry of J1529+2928 using the Apache Point Observatory
3.5m Telescope with the Agile frame-transfer camera. We obtained 45 s long exposures
with the BG40 filter over 1.7 h on UT 2015 April 10. We also obtained extensive
follow-up photometry using the McDonald 2.1m Telescope and the ProEM frame-transfer camera. 
We obtained 5-30 s long exposures with the BG40 (9.9 h), $i'$ (4.8 h), and $z'$ (7.0 h) 
filters on UT 2015 June 9-12 and Aug 12-14.

\subsection{Spectroscopy}

We used the 8m Gemini North telescope with the Gemini Multi-Object Spectrograph (GMOS)
to obtain medium resolution spectroscopy of J1529+2928 on 2015 July 7 and 10 as part of
the program GN-2015A-DD-8. 
We obtained a sequence of 26$\times$180 s exposures with the R831 grating and a 0.5$\arcsec$ slit, providing
wavelength coverage from 5500 \AA\ to 7600 \AA\ and a resolving power of 4396.
We also obtained a sequence of 20$\times$230 s exposures with the B1200 grating and a 0.5$\arcsec$ slit,
providing wavelength coverage from 3680 \AA\ to 5140 \AA\ and a resolving power of 3744.
Each spectrum has a comparison lamp exposure taken within 10 min of the observation time.
We measure radial velocities using the cross-correlation package RVSAO.

\section{Results}

\subsection{The Period}

\begin{figure}
\includegraphics[width=2.7in,angle=-90]{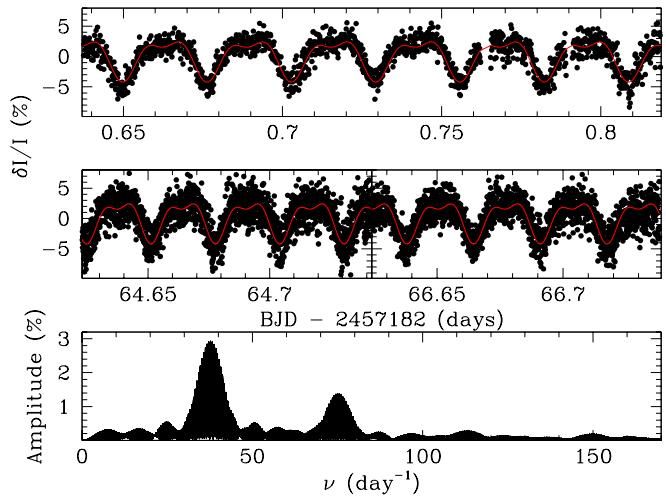}
\caption{High speed BG40-filter photometry of J1529+2928 over 9.9 h (top and middle panels). The
Fourier transform (bottom panel) shows a main peak at 37.74917(9) cycles d$^{-1}$ and its first
harmonic at 75.49834(21) cycles d$^{-1}$. The solid red line shows the predicted light curve based
on these two frequencies.}
\label{fig:lc}
\end{figure}

Figure \ref{fig:lc} shows the BG40-filter light curve of J1529+2928 obtained over 3 different nights
at the McDonald Observatory 2.1m telescope. There are only two significant peaks in the Fourier Transform;
the main peak is at 37.74917(9) cycles d$^{-1}$, 2288.792(6) s or 38.1 min, with a 2.95(4)\% amplitude,
and its first harmonic is detected at 75.49834(21) cycles d$^{-1}$. 
The light curve shows a strong dip in light that lasts about half the phase, a feature that
looks like an eclipse. There is no evidence of any change in the
`eclipse' times over the 66 day baseline of these observations. 

Our APO data with the BG40 filter span only 1.7 h, but the frequency and amplitude of
the variations are consistent with the McDonald data within the errors. We detect 3.19(7)\% photometric
variations with a period of 2250(10) s. Similarly, we detect variations with periods of 2289(1) s and
2288(13) s in the $i'$ and $z'$ data, respectively. Since the BG40 filter data from the McDonald 2.1m
telescope have the highest signal-to-noise ratio and the longest baseline, we adopt the best-fit period
of 2288.792(6) s for J1529+2928 for the remainder of the paper.

\subsection{No Stellar or Brown Dwarf Companions}

\begin{figure}
\includegraphics[width=1.8in]{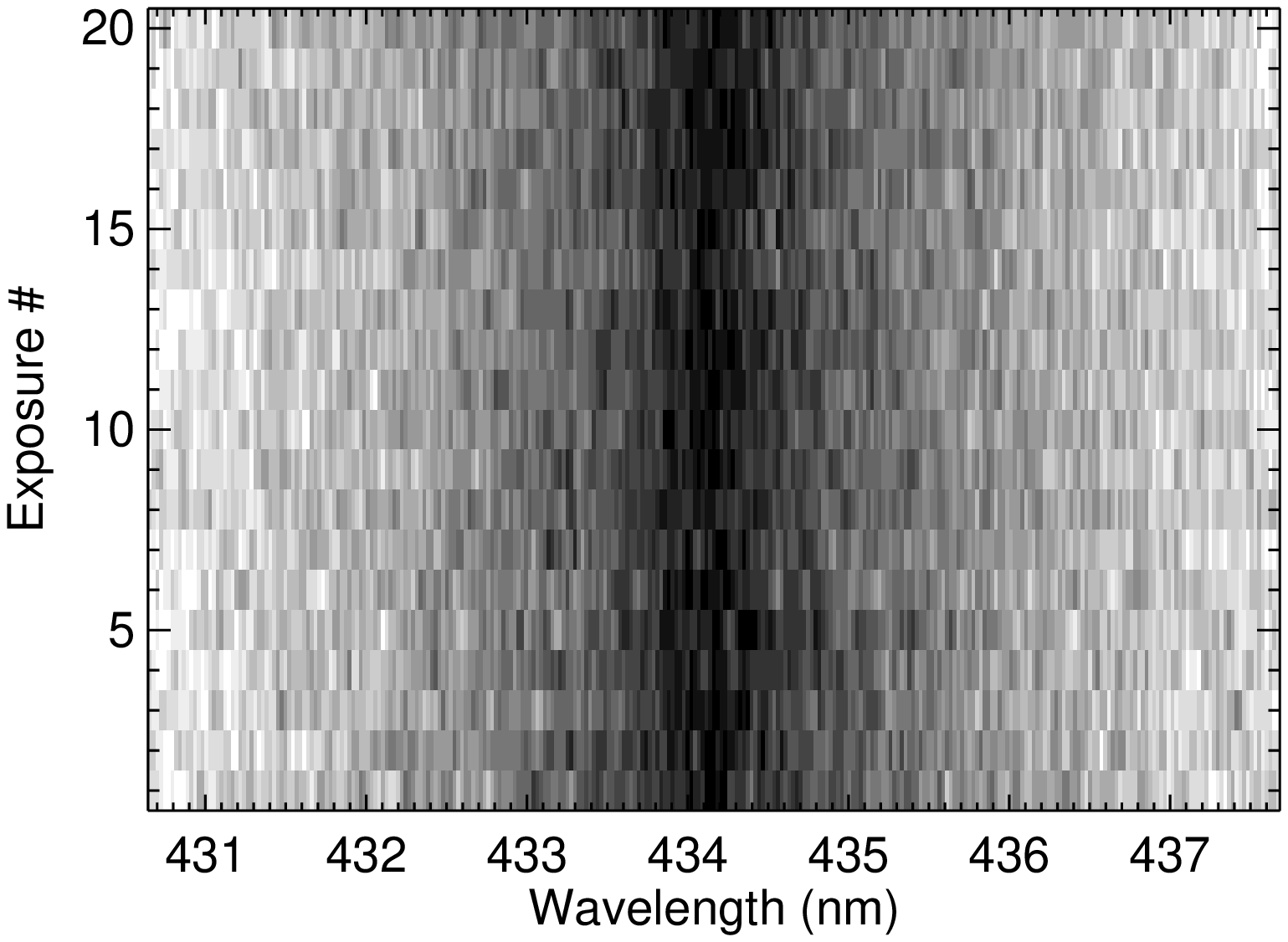}
\hspace{-0.5in}\includegraphics[width=1.8in]{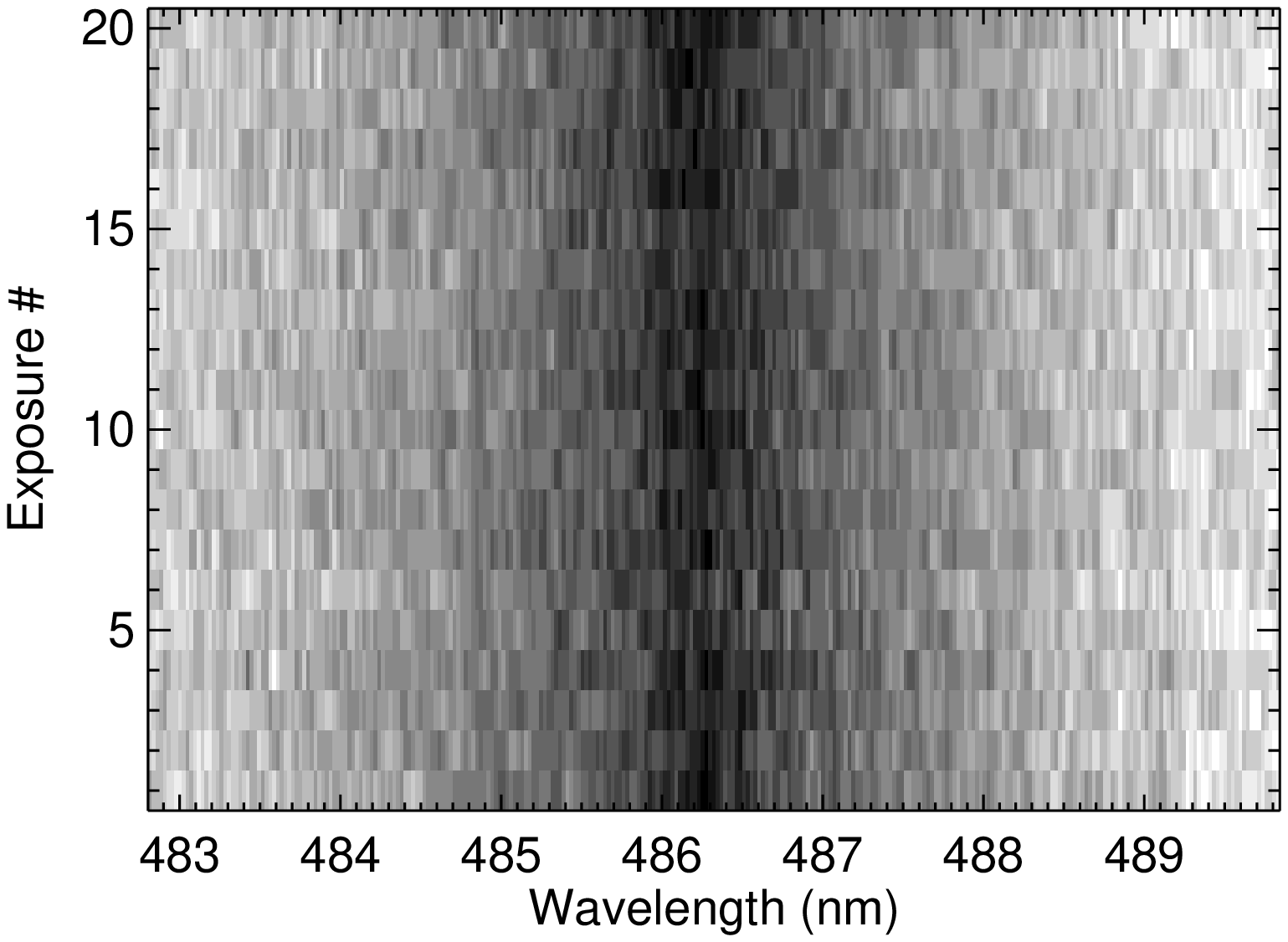}
\includegraphics[width=1.8in]{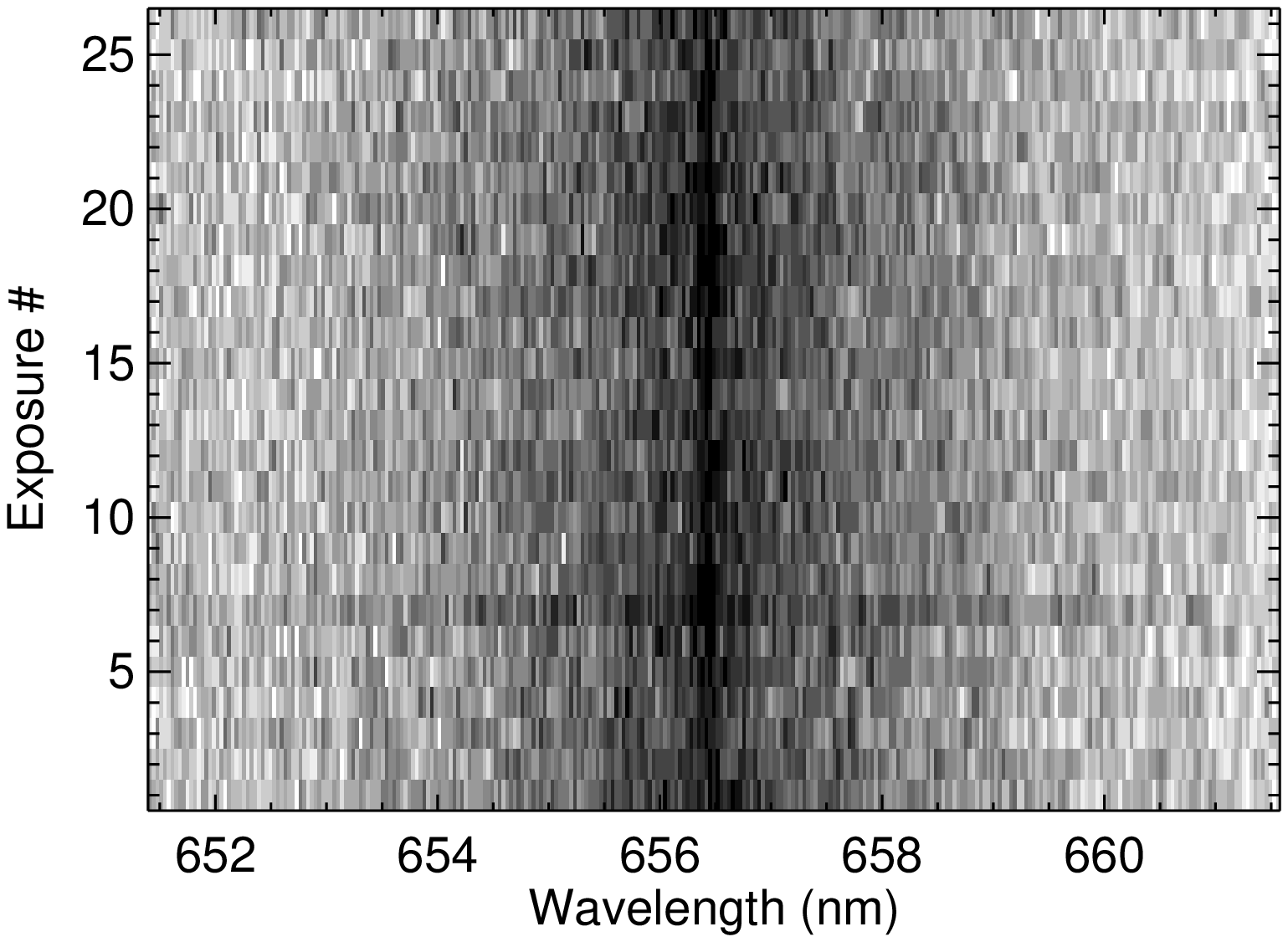}
\caption{Gemini time-resolved spectroscopy of H$\gamma$ (top left) and H$\beta$ (top right)
over 81 min, and that of H$\alpha$ (bottom left panel) over 85 min.}
\label{fig:gmos}
\end{figure}

Figure \ref{fig:gmos} shows trailed GMOS spectra of J1529+2928 for H$\gamma$, H$\beta$, and H$\alpha$.
The latter was observed with the R831 grating over 85 min, and the former two lines were observed with
the B1200 grating over 81 min. It is clear from this figure (especially given the narrow H$\alpha$ core) that
there is no evidence of significant velocity variations in any of the Balmer lines. Forcing a fit
at the photometric period of 2288.792 s yields a velocity semi-amplitude of $K = 3.6$ km s$^{-1}$. However,
the F-test p-value is 0.94, which implies that this fit is consistent with scatter around the mean velocity.
We perform 10,000 Monte-Carlo simulations assuming Gaussian errors and fixed period, and find that
the average velocity semi-amplitude is $K = 4.9 \pm 2.7$ km s$^{-1}$.

Figure \ref{fig:fit} shows our 1D atmospheric model fits to the SDSS and Gemini spectra of J1529+2928.
Fitting H$\beta$ through H8, we derive $T_{\rm eff} = 11450 \pm 200$ K and $\log{g} = 8.88 \pm 0.06$ from
the SDSS spectrum and $T_{\rm eff} = 11880 \pm 170$ K and $\log{g} = 8.78 \pm 0.04$ from the combined
Gemini spectrum. Since Gemini observations have higher resolution and signal-to-noise ratio, we adopt
the parameteres from the GMOS data. Applying the 3D model atmosphere corrections \citep{tremblay13} revise the
temperature and surface gravity to $T_{\rm eff} = 11580 \pm 170$ K and $\log{g} = 8.65 \pm 0.04$, respectively.
These correspond to $M = 1.02 \pm 0.03 M_{\odot}$, $M_g = 12.8$ mag, $d = 83 \pm 3$ pc, $R = 0.0079 R_{\odot}$,
$log{L/L_{\odot}} = -3.0$, and an age of 1.3 Gyr.

Given the photometric period and the lack of significant velocity variations, the mass function
is $f = 3 \pm 5 \times 10^{-7} M_{\odot}$. If the eclipses are due to a stellar or brown dwarf
companion, the inclination would have to be almost $90^{\circ}$. For an edge-on orbit, the companion
would be 0.007 $M_{\odot}$ , or 7 $M_{\rm Jupiter}$ at a separation of 0.38 $R_{\odot}$. Hence,
our radial velocity measurements rule out all stellar and brown dwarf companions around J1529+2928.
In addition, all solid-body planetary objects are also ruled out as they would lead to very short
transit durations of $<2$ min.

J1529+2928 has a systemic velocity of $\gamma = 39 \pm 10$ km s$^{-1}$ as measured from the SDSS
spectrum. The expected gravitational redshift is 81.8 km s$^{-1}$. Hence, the true systemic velocity
of J1529+2928 should be $-42.8 \pm 10$ km s$^{-1}$.

\section{Discussion}

\subsection{Not Pulsations}

\begin{figure}
\vspace{-1.3in}
\hspace{-0.5in}
\includegraphics[width=4.0in]{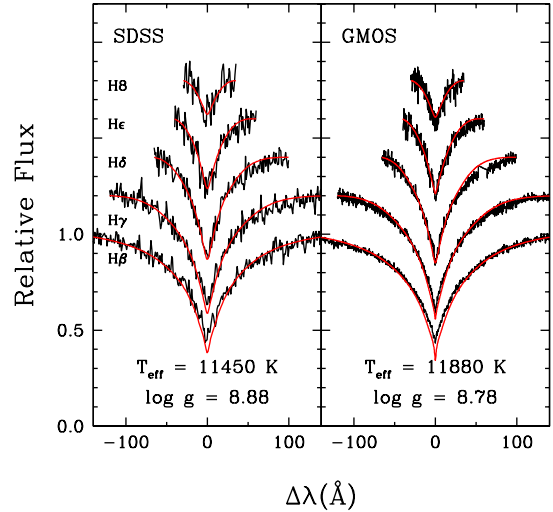}
\vspace{-1.3in}
\caption{Model fits (red lines) to the Balmer line profiles of J1529+2928 from the SDSS
(left panel) and Gemini (right panel) spectroscopy. Both fits show that H$\gamma$ and H$\beta$
are slightly filled in due to the flux contribution from a relatively cool starspot.}
\label{fig:fit}
\end{figure}

Our 1D atmospheric parameters for J1529+2928 are within 1$\sigma$ of the atmospheric parameters for
the massive pulsating ZZ Ceti BPM 37093 (see \S 2.1 and Fig. \ref{fig:fit}). However, the photometric dips
seen in J1529+2928 cannot be due to pulsations. All known DAVs have pulsation periods of $\leq$ 1200 s
\citep[e.g.,][]{mukadam06}. BPM 37093 and GD 518, the two pulsators with $M\geq1M_{\odot}$, have pulsation
periods of 512-635 s and 425-595 s, respectively \citep{metcalfe04,hermes13}. 
On the theoretical side, the buoyancy cutoff period for typical-mass WDs is $\sim$1000 s, and almost
certainly $<2000$ s \citep{hansen85}. Hence, the 38 min period variability in J1529+2928 is not due to
pulsations.

\subsection{Not a Disintegrating Planet}

The Roche Limit for Earth density objects around J1529+2928 is $1.4 R_{\odot}$ \citep{agol11}. 
Hence, a putative 38 min period orbiting planet would be tidally disrupted.
Such a planet would be falling apart after hundreds of orbits, yet Figure \ref{fig:lc} clearly
demonstrates that the photometric dips are constant over several weeks/months and that the hypothetical
object is not appreciably changing. This implies that the `eclipses' are not from a transiting object,
but rather from a surface feature on the white dwarf itself.

Unlike the candidate disintegrating planet around WD 1145+017 \citep{vanderburg15}, there is
no evidence of a debris disk around J1529+2928 based on the 3.6 $\mu$m photometry from the
Wide-field Infrared Survey Explorer \citep[WISE,][]{wright10} mission.
In addition, a relatively cool eclipsing planetary companion would lead to increasing eclipse
depth as a function of wavelength. This is not seen in our photometric observations. Hence, the source
of the photometric dips in J1529+2928 is not a disintegrating planet.

\subsection{A Dark Spot on J1529+2928}

\begin{figure}
\includegraphics[width=2.7in,angle=-90]{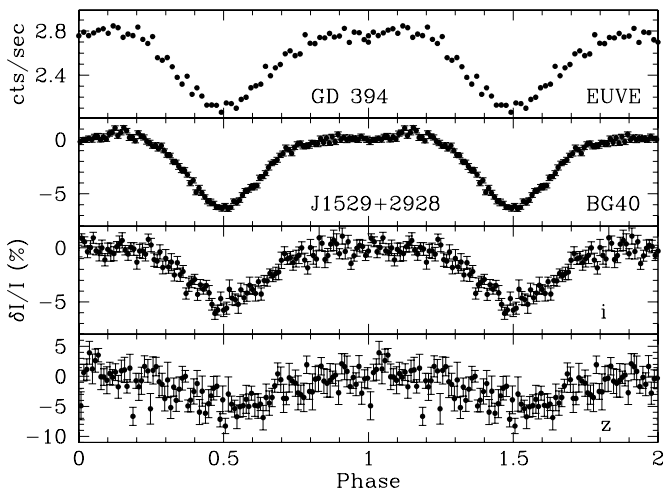}
\caption{EUVE light curve of GD 394 \citep[top panel,][]{dupuis00}, and
the phased and binned light curves of J1529+2928 in the BG40, $i'$ and $z'$ filters.
We used the best-fit period of 2288.792 s from the BG40 light curve to phase the $i'$ and $z'$ filter
data.}
\label{fig:color}
\end{figure}

Figure \ref{fig:color} shows the phased and binned BG40, $i'$, and $z'$ light curves of J1529+2928.
Here we used the best-fit period from the BG40 data, 2288.792 s, to phase the data in the other filters.
The semi-amplitude of the photometric dips change from $2.95 \pm 0.04$\% in the BG40 filter, to $2.30 \pm 0.11$\% and $2.55 \pm 0.32$\%
in the $i'$ and $z'$ filter data, respectively. The $z'$ data are relatively noisy and
do not provide strong constraints on this system. However, the dips in the
$i'$ filter are significantly shallower than those in the BG40 filter. This clearly rules out a cool substellar
companion around J1529+2928, and instead favors a starspot on the surface of the white dwarf.
The spots on BOKS 53856 and WD 2359-434 were stable over several months \citep{holberg11,gary13},
though the latter shows a slight decrease in amplitude over two years (B. Gary 2015, private communication). 
Hence, starspots can explain the stability of the photometric dips in our data spanning four months. 

Figure \ref{fig:color} also shows the Extreme Ultraviolet Explorer (EUVE) data on GD 394
from \citet{dupuis00}. The similarities between the GD 394 and J1529+2928 light curves are
stunning. GD 394 is a relatively hot white dwarf with $T_{\rm eff} = 39,440 \pm 350$ K
and $\log{g} = 7.90 \pm 0.07$. The optical and ultraviolet spectra show strong Si features
as well as a variety of trace elements including C, N, O, Al, and Fe. The Si abundance
by number is (Si/H) $\sim 10^{-5}$. \citet{dupuis00} suggest that the 1.15 d period variability seen
in the EUVE light curve of GD 394 is due to inhomogeneous abundance distribution of metals over the
surface of the star. In their model, an almost circular spot with a higher heavy element abundance
reduces the EUV emission over the spot region, causing 25\% dips in brightness.

We do not have strong constraints on the metal abundances in J1529+2928 given the lack of high-resolution
optical spectroscopy or ultraviolet data. However, there is a weak Ca K line in our GMOS spectrum at the
same velocity as the Balmer lines. We measure an abundance of $\log{(Ca/H)} = -7.91$ from this line, but the 
spectrum is relatively noisy in this wavelength range and we take this measurement as
an upper limit on the Ca abundance. Hence, metals are likely present in the atmosphere of J1529+2928
and are the likely cause of the periodic dips seen in its light curve. Since the opacity due to metals
essentially decreases with increasing wavelength, a dark spot with a higher heavy element abundance
would naturally explain the decrease in `eclipse' depth in the BG40 and $i-$band data.

Both the SDSS and Gemini spectra (see Fig. \ref{fig:fit}) show that the observed line profiles are
shallower than predicted by the best-fit model.
The same is also true for the H$\alpha$ line profiles. Assuming an inclination of $90^{\circ}$,
a 10,000 K spot that covers 14\% of the surface area of the white dwarf would explain the observed
dips in both BG40 and $i'$ filters. Due to the slight temperature difference between the spot
and the white dwarf, such a spot also helps explain the observed H$\beta$ and H$\alpha$ line profiles.
However, there are significant degeneracies in modelling
of starspots due to unknown inclination, number of spots, spot latitude, and temperature \citep{dupuis00},
and further observations will be necessary to constrain the temperature of the spot region more accurately.

It is difficult to explain the presence of a spot on a white dwarf without invoking a magnetic field.
\citet{dupuis00} constrain the magnetic field in GD 394 to $B<12$ kG and they argue that such low field
strengths may be enough to channel the accreted metals onto a spot if the accretion rate is low.
Based on the lack of detection of Zeeman splitting of the narrow H$\alpha$ core, we derive
an upper limit of $B<70$ kG. \citet{maxted00}, \citet{wickram00}, and \citet{brinkworth05} demonstrate
that spotlike magnetic field enhancements extend to low field stars. They find a surface-wide 70 kG field
and a localized 500 kG field to explain the 1.44 d period 2\% peak-to-peak amplitude variations seen in
WD 1953-011. Hence, a similar field structure could potentially explain the presence of spots on GD 394
and J1529+2928.

\subsection{Implications for the Kepler/K2 missions and the LSST}

GD 394 and J1529+2928 present excellent examples of apparently non-magnetic and isolated white dwarfs
with spots. Both stars show significant dips in their light curves on hour-to-day timescales that are
relatively easy to detect.
The discovery of low-level (up to 0.2\%) hour-to-day timescale variations in the Kepler
data on apparently non-magnetic and isolated white dwarfs by \citet{maoz15} is therefore quite interesting.
\citet{maoz15} present several scenarios to explain their observations, including but not limited to magnetic
spots, hot spots from an interstellar medium accretion flow, transits by small planetary companions, or
the reflection effect on a brown dwarf or giant planet companion. 

Based on the detection of hour/day
timescale photometric variations in strongly magnetic (e.g., RE J0317-853), weakly magnetic
(e.g., WD 2359-434), and apparently non-magnetic (e.g., GD 394 and J1529+2928) white dwarfs,
we suggest that many of the variable white dwarfs in the Kepler and K2 mission likely have
starspots due to surface inhomogenities and that the observed variability is a direct indicator
of the rotation periods of these white dwarfs.
The K2 mission will observe several hundred white dwarfs, and even with its
decreased sensitivity compared to the original Kepler mission, it should still be able to find
a relatively large number of variable white dwarfs and measure their rotation rates.

The LSST will image 13 million white dwarfs down to $r=24.5$ mag and the design goals include
photometric repeatability of 5 mma at the bright end. The photometric precision of each visit
will be worse at fainter magnitudes, e.g. 1\% at $r=21$ mag and 2\% at $r=22$ mag \citep{ivezic08}. 
Only one of the 15 white dwarfs (6.7\%) observed by the Kepler mission, BOKS 53856, displays photometric
variations larger than a few per cent. Eleven out of the 74 (15\%) white dwarfs within 20 pc of the Sun are
magnetic with $B\geq$ 100 kG \citep{kawka07}. In addition, \citet{brinkworth13} find 1-2\% variability in 2/3 of
their $B\geq 100$ kG magnetic white dwarf sample, implying an overall variability fraction of 10\%. 
\citet{sion14} find the fraction of magnetic white dwarfs in the 25 pc local sample to be
$\geq8$\%, but \citet{liebert03} demonstrate that the fraction could be substantially higher than 10\% for
$B>10$ kG. Hence, with a 1\% precision at $r=21$ mag, the LSST may find $\sim 10^5$ white dwarfs that
show variations on hour/day timescales due to the presence of magnetic fields.

\section{Conclusions}

We present the discovery of 38 min period eclipse-like events in the light curve of the massive white
dwarf J1529+2928. We rule out stellar and brown dwarf companions based on the lack of significant
radial velocity variations, and demonstrate that the dips in the light curve are likely caused by
a dark spot that comes into view every 38 min due to the rotation of the white dwarf. 

The presence of a spot on J1529+2928 almost certainly requires a magnetic field.
However, there is no evidence of Zeeman splitting of the Balmer lines in our optical spectroscopy,
constraining the magnetic field strength to $B<70$ kG. Follow-up high resolution optical 
spectroscopy or spectropolarimetry would be useful to search for a magnetic field
and constrain its strength. 

A weak Ca K line is present in our spectrum, though follow-up UV spectroscopy would be useful
to confirm the presence of metals and constrain their abundances in J1529+2928. \citet{dupuis00}
demonstrate that accreted metals concentrated on a spot by a magnetic field would lead to
an enhanced opacity source when the spot is in view, and this could explain the
observed EUV light curve of GD 394. The same model would also explain the differing
`eclipse' depths in J1529+2928 in different filters. \citet{dupuis00} predict that the UV
and optical light curves should be in antiphase due to the flux redistribution effect.
Follow-up concurrent UV and optical photometry observations can test this hypothesis.
 
Finally, based on the discovery of significant photometric variations in apparently non-magnetic white
dwarfs like GD 394 and J1529+2928, we discuss the high incidence of photometric variability observed
in the Kepler and K2 missions. We argue that the source of the variability is most likely related
to the presence of weak magnetic fields, and that current and future transient surveys like the LSST
should find a significant number of white dwarfs that display hour-to-day timescale photometric variations.

\acknowledgements
We gratefully acknowledge the support of the NSF and NASA under grants AST-1312678, AST-1312983,
and NNX14AF65G. 

{\it Facilities:} \facility{Gemini (GMOS), APO 3.5m (Agile), McDonald 2.1m (ProEM)}


\begin{thebibliography}

\bibitem[Agol(2011)]{agol11} Agol, E.\ 2011, \apjl, 731, L31
\bibitem[Aznar Cuadrado et al.(2004)]{aznar04} Aznar Cuadrado, R., Jordan, S., Napiwotzki, R., et al.\ 2004, \aap, 423, 1081 
\bibitem[Barstow et al.(1995)]{barstow95} Barstow, M.~A., Jordan, S., O'Donoghue, D., et al.\ 1995, \mnras, 277, 971 
\bibitem[Brinkworth et al.(2005)]{brinkworth05} Brinkworth, C.~S., Marsh, T.~R., Morales-Rueda, L., et al.\ 2005, \mnras, 357, 333 
\bibitem[Brinkworth et al.(2013)]{brinkworth13} Brinkworth, C.~S., Burleigh, M.~R., Lawrie, K., Marsh, T.~R., \& Knigge, C.\ 2013, \apj, 773, 47 
\bibitem[Dupuis et al.(2000)]{dupuis00} Dupuis, J., Chayer, P., Vennes, S., Christian, D.~J., \& Kruk, J.~W.\ 2000, \apj, 537, 977 
\bibitem[Ferrario et al.(1997)]{ferrario97} Ferrario, L., Vennes, S., Wickramasinghe, D.~T., Bailey, J.~A., \& Christian, D.~J.\ 1997, \mnras, 292, 205 
\bibitem[Gary et al.(2013)]{gary13} Gary, B.~L., Tan, T.~G., Curtis, I., Tristram, P.~J., \& Fukui, A.\ 2013, Society for Astronomical Sciences Annual Symposium, 32, 71 
\bibitem[Gianninas et al.(2011)]{gianninas11} Gianninas, A., Bergeron, P., \& Ruiz, M.~T.\ 2011, \apj, 743, 138 
\bibitem[Hallakoun et al.(2015)]{hallakoun15} Hallakoun, N., Maoz, D., Kilic, M., et al.\ 2015, \mnras, submitted, arXiv:1507.06311 
\bibitem[Hansen et al.(1985)]{hansen85} Hansen, C.~J., Winget, D.~E., \& Kawaler, S.~D.\ 1985, \apj, 297, 544 
\bibitem[Hermes et al.(2013)]{hermes13} Hermes, J.~J., Kepler, S.~O., Castanheira, B.~G., et al.\ 2013, \apjl, 771, L2 
\bibitem[Hermes et al.(2014)]{hermes14} Hermes, J.~J., Brown, W.~R., Kilic, M., et al.\ 2014, \apj, 792, 39 
\bibitem[Holberg \& Howell(2011)]{holberg11} Holberg, J.~B., \& Howell, S.~B.\ 2011, \aj, 142, 62 
\bibitem[Ivezic et al.(2008)]{ivezic08} Ivezic, Z., Tyson, J.~A., Abel, B., et al.\ 2008, arXiv:0805.2366 
\bibitem[Kawka et al.(2007)]{kawka07} Kawka, A., Vennes, S., Schmidt, G.~D., Wickramasinghe, D.~T., \& Koch, R.\ 2007, \apj, 654, 499 
\bibitem[Kilic et al.(2011)]{kilic11} Kilic, M., Brown, W.~R., Kenyon, S.~J., et al.\ 2011, \mnras, 413, L101 
\bibitem[Kleinman et al.(2013)]{kleinman13} Kleinman, S.~J., Kepler, S.~O., Koester, D., et al.\ 2013, \apjs, 204, 5 
\bibitem[Landolt(1968)]{landolt68} Landolt, A.~U.\ 1968, \apj, 153, 151 
\bibitem[Liebert et al.(2003)]{liebert03} Liebert, J., Bergeron, P., \& Holberg, J.~B.\ 2003, \aj, 125, 348 
\bibitem[Loeb \& Gaudi(2003)]{loeb03} Loeb, A., \& Gaudi, B.~S.\ 2003, \apjl, 588, L117 
\bibitem[Maoz et al.(2015)]{maoz15} Maoz, D., Mazeh, T., \& McQuillan, A.\ 2015, \mnras, 447, 1749
\bibitem[Maxted et al.(2000)]{maxted00} Maxted, P.~F.~L., Ferrario, L., Marsh, T.~R., \& Wickramasinghe, D.~T.\ 2000, \mnras, 315, L41 
\bibitem[Metcalfe et al.(2004)]{metcalfe04} Metcalfe, T.~S., Montgomery, M.~H., \& Kanaan, A.\ 2004, \apjl, 605, L133 
\bibitem[Mukadam et al.(2004)]{mukadam04} Mukadam, A.~S., Mullally, F., Nather, R.~E., et al.\ 2004, \apj, 607, 982 
\bibitem[Mukadam et al.(2006)]{mukadam06} Mukadam, A.~S., Montgomery, M.~H., Winget, D.~E., Kepler, S.~O., \& Clemens, J.~C.\ 2006, \apj, 640, 956 
\bibitem[Parsons et al.(2013)]{parsons13} Parsons, S.~G., G{\"a}nsicke, B.~T., Marsh, T.~R., et al.\ 2013, \mnras, 429, 256 
\bibitem[Shporer et al.(2010)]{shporer10} Shporer, A., Kaplan, D.~L., Steinfadt, J.~D.~R., et al.\ 2010, \apjl, 725, L200
\bibitem[Sion et al.(2014)]{sion14} Sion, E.~M., Holberg, J.~B., Oswalt, T.~D., et al.\ 2014, \aj, 147, 129 
\bibitem[Tremblay et al.(2013)]{tremblay13} Tremblay, P.-E., Ludwig, H.-G., Steffen, M., \& Freytag, B.\ 2013, \aap, 559, A104 
\bibitem[Vanderburg et al.(2015)]{vanderburg15} Vanderburg, A., Johnson, J. A., Rappaport, S., et al. 2015, \nat, in press
\bibitem[Wickramasinghe \& Ferrario(2000)]{wickram00} Wickramasinghe, D.~T., \& Ferrario, L.\ 2000, \pasp, 112, 873
\bibitem[Wright et al.(2010)]{wright10} Wright, E.~L., Eisenhardt, P.~R.~M., Mainzer, A.~K., et al.\ 2010, \aj, 140, 1868
\bibitem[Zucker et al.(2007)]{zucker07} Zucker, S., Mazeh, T., \& Alexander, T.\ 2007, \apj, 670, 1326 

\end{thebibliography}
\end{document}